\documentclass[reprint,
superscriptaddress,
amsmath,
amssymb,
prb
]{revtex4-2} %

\setcitestyle{numbers,sort&compress,round}

\usepackage{graphicx}
\usepackage{physics}
\usepackage{color}
\usepackage{braket}
\usepackage[colorlinks=true, citecolor=blue, linkcolor=blue, urlcolor=blue]{hyperref}
\usepackage{comment}
\usepackage{xcolor}
\usepackage[normalem]{ulem}
\usepackage{xspace}

\usepackage{amsmath} %
\usepackage{bm}  %

\newcommand{\e}{$e$}
\newcommand{\kk}{\bm{k}}
\renewcommand{\qq}{\bm{q}}

\hypersetup{
     colorlinks   = true,
     linkcolor    = blue,
     citecolor    = blue,
     urlcolor     = blue,
}
\definecolor{mypink}{rgb}{0.858,0.188,0.478}

\newcommand{\autocite}[1]{\cite{#1}}

\makeatletter

\AtBeginDocument{

  \setlength{\abovedisplayskip}{6pt}

  \setlength{\belowdisplayskip}{6pt}

  \setlength{\abovedisplayshortskip}{3pt}

  \setlength{\belowdisplayshortskip}{3pt}

}

\begin{document}
\title{A Compact Tensor Representation of
Electron-Phonon Interactions} %

\author{Yao Luo}%
\affiliation{Department of Applied Physics and Materials Science, and Department of Physics, California Institute of Technology, Pasadena, California 91125, USA}

\author{Shiv S. Seshan}%
\affiliation{Department of Applied Physics and Materials Science, and Department of Physics, California Institute of Technology, Pasadena, California 91125, USA}

\author{Khoa Le}%
\affiliation{Department of Applied Physics and Materials Science, and Department of Physics, California Institute of Technology, Pasadena, California 91125, USA}

\author{Marco Bernardi}
\affiliation{Department of Applied Physics and Materials Science, and Department of Physics, California Institute of Technology, Pasadena, California 91125, USA}
\email{bmarco@caltech.edu}

\begin{abstract}

Understanding microscopic interactions in materials is central to advances in semiconductors, energy, and quantum technologies. 
Yet the large size of first-principles interaction datasets remains a bottleneck, limiting predictive calculations of physical properties and dynamics in complex materials. 
Here, we overcome this challenge with a compact tensor representation of electron–phonon (\e-ph) interactions that preserves momentum conservation and Hermiticity, achieving gauge-invariant compression and linear scaling with system size and momentum grids.  
Across diverse materials, it compresses \e-ph interactions by two to three orders of magnitude with a minimal loss in accuracy.    
Retaining a tiny fraction ($\sim$0.1\%) of the data enables quantitative predictions of electronic scattering, transport, polarons, and superconductivity, with speedups approaching three orders of magnitude. 
The framework extends naturally to exciton–phonon interactions, where matrix-based decomposition fails. These results reveal an efficient low-rank representation of microscopic interactions and establish a systematically tunable, scalable formalism for high-throughput and many-body calculations.

\end{abstract}
\maketitle 
\vspace{100pt}
\section*{Introduction}
\vspace{-10pt}
Advances in first-principles calculations of electron–phonon (\e-ph) interactions have enabled quantitative predictions of charge and spin transport, band renormalization, polaron formation, and phonon-mediated superconductivity across a wide range of materials~\cite{Bernardi2014,Li2015,Zhou2016,Liu2017,Jhalani2017,Ma2018,Zhou2018,Sohier2018,Zhou2019,Samuel2020,Lee2020_twophonon,Desai2021,Chang2022,Park_2022,Zheng2023,luo_first-principles_2025}. These calculations use density functional theory (DFT) and density functional perturbation theory (DFPT)\cite{Giannozzi2001} to obtain the \e-ph interaction matrices~\cite{Perturbo2021}, $g_{mn,\nu}(\kk,\qq)$, describing scattering between electronic states $\ket{n\kk}$ and $\ket{m\kk+\qq}$ via a phonon mode $(\nu,\qq)$. Unlike model Hamiltonians, which describe interactions using a small number of coupling constants, first-principles calculations generate \e-ph couplings for every combination of electronic states and phonon modes. This microscopic resolution is essential for predictive accuracy, but it creates a formidable data challenge as the resulting datasets can contain billions of entries, making their computation, storage, and interpolation a major bottleneck. %
\\
\indent 
Dimensionality reduction offers a data-driven route to preserving accuracy while reducing model complexity. Accordingly, low-rank representations have become central in condensed matter physics, quantum science, and electronic structure theory, including tensor-network representations of many-body wave functions~\cite{rmp-tns,tns-2}, tensor trains for high-order Feynman diagrams~\cite{TT-feynman-2022,TT-impurity-2023,TT-pa-2025}, tensor hypercontraction~\cite{thc-1,thc-2,thc-3} and density fitting~\cite{df-qchem,Yang-df-2026} in quantum chemistry, and feature compression for atomistic machine learning~\cite{Ceriotti2018,Ceriotti2021,Darby2022}.
%
For microscopic interactions in materials, we recently introduced singular-value decomposition (SVD) to compress \e-ph matrices, and tensor decomposition to reveal low-rank structure in phonon–phonon interactions~\cite{SVD-eph-prx2024,luo-pcp-prl2025}. However, \e-ph interactions are still represented and compressed as matrices, leaving their underlying tensor structure and low-rank representations unexplored.
\\
\indent
Here, we introduce a momentum-conserving CANDECOMP/PARAFAC (MCP) decomposition of first-principles \e-ph interactions. MCP treats the coupling as an order-three tensor whose slots correspond to the initial electronic state, final electronic state, and phonon mode. This formulation compresses electronic and phonon momenta together with band and mode indices, achieves gauge-independent compression by transforming covariantly under electronic gauge rotations, and enforces crystal-momentum conservation and Hermiticity in the low-rank ansatz. 
Across six systems (LiF, Si, GaAs, MoS$_2$, La$_2$CuO$_4$, and large-angle twisted bilayer graphene (TBG)), MCP achieves \e-ph compression factors of $10^2$–$10^3$, substantially outperforming SVD while retaining quantitative accuracy for relaxation times, transport, superconductivity, and polarons. 
MCP accelerates \e-ph calculations by similar factors, approaching the cost of model Hamiltonians. 
Its advantage is even more pronounced for exciton–phonon interactions: while matrix-based SVD yields negligible compression, MCP reveals a compact low-rank tensor structure in monolayer CrBr$_3$. These results establish MCP as a generic and scalable tensor representation of microscopic interactions in materials, accelerating state-of-the-art first-principles workflows~\cite{Perturbo2021, epw-2016} and extending the reach of quantitative many-body simulations to increasingly complex materials and interactions~\cite{luo_first-principles_2025}.

\vspace{-14pt}
\section*{Results}
\vspace{-6pt}
\subsection*{Tensor compression of \e-ph interactions}
\begin{figure}
    \centering
    \includegraphics[width=0.96\linewidth]{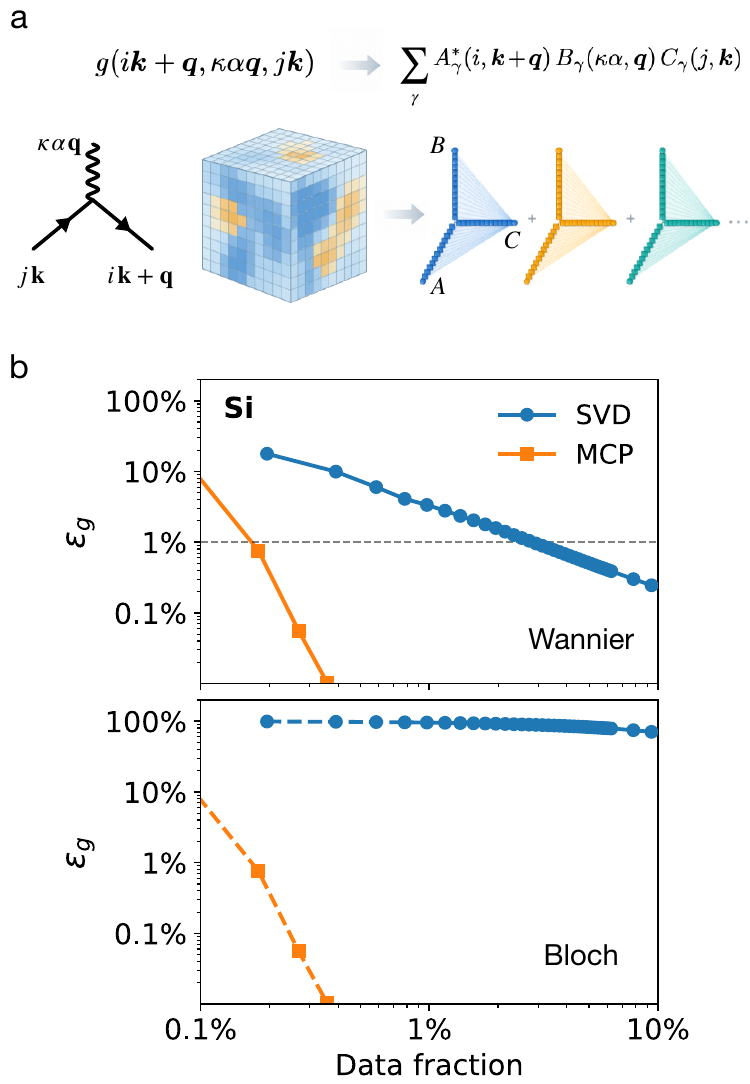}
    \caption{(a) Schematic of MCP decomposition of \e-ph interactions. (b) Reconstruction error $\epsilon_{g}$ for \e-ph interactions in Si, shown as a function of the retained data fraction $p$ in Eq.~\eqref{eq:datafrac}, comparing MCP and SVD compression methods using a smooth Wannier gauge (top) and the Bloch gauge from DFT (bottom). The dashed line indicates $\epsilon_g = $ 1\%.}
    \label{fig:gauge_comparison}
\end{figure}

\vspace{-8pt}
The central idea is to recast the \e-ph interaction as a tensor. Let $g_{ij,\kappa\alpha}(\kk,\qq)$ denote the \e-ph coupling in a basis of electronic orbitals $i,j$ and atomic displacements $\kappa\alpha$, with $\kappa$ labeling atoms and $\alpha$ Cartesian directions. By pairing each basis index with its associated momentum, we obtain\vspace{-6pt}
\begin{equation}
\label{eq:gtensor}
g_{ij,\kappa\alpha}(\kk,\qq) = g(i\kk+\qq,\kappa\alpha\qq,j\kk),
\end{equation}
whose three slots represent the final electronic state, atomic displacement, and initial electronic state. This form makes explicit the intrinsic three-particle structure and momentum conservation of \e-ph coupling~\cite{Ziman2001}.
\\
\indent
Originally developed for multiway data~\cite{Hitchcock1927,Carroll1970,harshman1970}, the CP decomposition is now widely used for tensor-data compression in machine learning~\cite{lebedev2015speedingupconvolutionalneuralnetworks,chen2022TensoRF,pmlr-v80-lacroix18a}. We introduce a momentum-conserving CP decomposition for first-principles \e-ph interactions through the MCP ansatz
\begin{equation}
 \tilde{g}_{ij,\kappa\alpha}(\kk,\qq)
 =\sum_{\gamma=1}^{R_{\rm MCP}}
 A_\gamma^*(i,\kk+\qq)\,
   B_\gamma(\kappa\alpha,\qq)\,
   C_\gamma(j,\kk)
   +\mathrm{H.c.},
 \label{eq:mcp}
\end{equation}
where $R_{\rm MCP}$ is the tensor rank and $\mathrm{H.c.}$ denotes the Hermitian conjugate. 
MCP approximates the order-three tensor $g$ as a sum of order-one factors: electronic modes $A_\gamma$ and $C_\gamma$ for the final and initial states, and a vibrational mode $B_\gamma$, as illustrated in
Fig.~\ref{fig:gauge_comparison}(a). 
This yields a compact form in which each term defines an \e-ph scattering channel in learned coordinates. The MCP decomposition preserves the physical structure of the interaction: the coupled momenta $(\kk+\qq,\qq,\kk)$ enforce crystal-momentum conservation, the Hermitian-conjugate term ensures Hermiticity, and the decomposition transforms covariantly under electronic gauge rotations.
%
\\
\indent
We obtain the MCP factors $\{\mathbf{A},\mathbf{B},\mathbf{C}\}$ by minimizing the reconstruction error of the DFPT interaction tensor on the coarse grid,
\begin{equation}\label{eq:mcp-loss}
    L[\mathbf{A},\mathbf{B},\mathbf{C}]
    =\sum_{\kk ij,\qq\nu}
    \alpha_{\nu\qq}
    \left\| g_{ij,\nu}(\kk,\qq)
      - \tilde{g}_{ij,\nu}(\kk,\qq) \right\|^2,
\end{equation}
where $\tilde g$ is the MCP ansatz in Eq.~\eqref{eq:mcp} in a phonon-mode basis and $\alpha_{\nu\qq}$ are mode-dependent weights. 
The MCP factors are determined using gradient-based optimization in \textsc{PyTorch}~\cite{pytorch}.
We characterize the compression via the retained data fraction $p$, defined as
\begin{equation}
\label{eq:datafrac}
p = \frac{\#\bm{A}+\#\bm{B}+\#\bm{C}}{\#\bm{g}}
  = \frac{R_{\mathrm{MCP}}
  (2N_bN_{k_c}+3N_aN_{q_c})}
  {3N_aN_b^2N_{k_c}N_{q_c}},
\end{equation}
where $\#\bm{X}$ denotes the number of stored elements in $\bm{X}$, $N_{k_c}$ and $N_{q_c}$ are the numbers of coarse-grid electron and phonon momenta, respectively, while $N_a$ and $N_b$ are the numbers of atoms and electronic bands. Smaller $p$ values correspond to stronger compression. The accuracy is quantified by the reconstruction error
\begin{equation}
\epsilon_g = \|\tilde{\bm{g}}-\bm{g}\|_F^2 \,\big/
\,\|\bm{g}\|_F^2,
\label{eq:rel-loss}
\end{equation}
evaluated in the electronic-band and phonon-mode basis.
The first-principles electronic structure and phonons are obtained from DFT and DFPT with \textsc{Quantum Espresso}~\cite{QE2009}; we use \textsc{Perturbo}~\cite{Perturbo2021} to compute the \e-ph couplings in the Wannier~\cite{Marzari2014} and Bloch gauges. See Methods and Supplemental Material (SM)~\cite{supplemental} for additional details.

The top panel of Fig.~\ref{fig:gauge_comparison}(b) compares MCP and SVD compression~\cite{SVD-eph-prx2024} for Si in the smooth Wannier gauge~\cite{Marzari2014}. MCP achieves a small error $\epsilon_g\!<\!1\%$ using only less than $0.2\%$ of the interaction data, corresponding to a compression factor greater than 500. Increasing the accuracy to $\epsilon_g\approx0.1\%$ requires only a modest increase to $\sim$0.3$\%$ of retained data. In contrast, SVD reaches $1\%$ error only after retaining approximately $3\%$ of the data. 
The order-of-magnitude gain in efficiency reflects the greater flexibility of the tensor representation: MCP compresses electronic bands, vibrational modes, and momenta jointly, whereas SVD treats the interaction as a collection of matrices and leaves the internal electronic and vibrational indices uncompressed.
\\
\indent
Another advantage of MCP is its gauge covariance. Repeating the compression analysis in the original DFT Bloch gauge [Fig.~\ref{fig:gauge_comparison}(b), bottom panel] leaves the MCP error curve unchanged, showing that its performance is independent of electronic gauge. In striking contrast, SVD provides no useful compression in the Bloch gauge $-$ its efficiency relies on rapid singular-value decay, which is linked to the smooth momentum dependence of $g(\kk,\qq)$ in the Wannier gauge~\cite{Griebel2018}. This smoothness is absent in the original Bloch gauge because the phases of the Bloch wave functions from DFT are set randomly across momentum space, leading to slow singular-value decay and poor SVD compression.
\\
\indent
The gauge covariance of the MCP ansatz guarantees identical accuracy under arbitrary unitary rotations of the electronic states. For a gauge transformation $U(\kk)$ in the electronic subspace~\cite{Wannier-rmp-2012}, the \e-ph matrices transform as $g_\nu(\kk,\qq)\rightarrow U^\dagger(\kk+\qq)g_\nu(\kk,\qq)U(\kk)$~\cite{Perturbo2021}. 
In the MCP ansatz, this transformation preserves both the form and rank of the decomposition: the lattice factor $B_\gamma$ is unchanged, and the electronic factors $A_\gamma$ and $C_\gamma$ transform \mbox{covariantly} (the same relation holds for $A_\gamma$ and $C_\gamma$): 
%
\begin{equation}
\label{eq:mcp-rotation}
A_{\gamma}(n\kk) = \sum_i U_{in}(\kk)\,A_{\gamma}(i\kk).
\end{equation}
\begin{figure}[!t]
    \centering
    \includegraphics[width=1.02\linewidth]{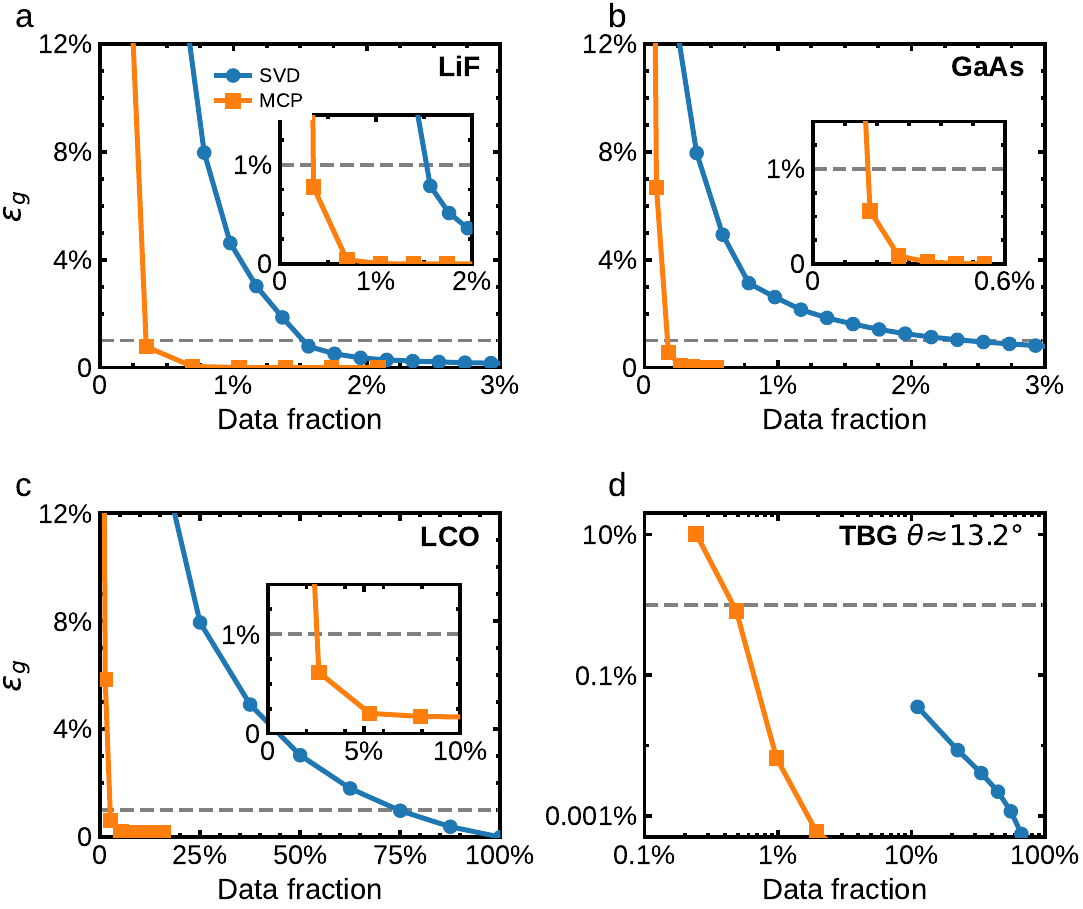}
    \caption{Reconstruction error $\epsilon_{g}$ for (a) LiF, (b) GaAs, (c) LCO, and (d) TBG, comparing SVD (blue; Wannier gauge) and MCP decomposition (orange), shown as a function of the retained data fraction $p$.
    The insets in (a)-(c) show a zoomed-in view for small data fractions, while panel (d) uses a log-log scale. The dashed line indicates the threshold $\epsilon_{g} = 1\%$.}

    \label{fig:compression_comparison}
\end{figure}
\noindent
Therefore, every rank-$R_{\mathrm{MCP}}$ representation in one gauge maps onto a representation of the same rank in any other gauge, providing gauge-invariant MCP compression~\cite{supplemental}.
\\
\indent
Across materials with diverse electronic structures, bonding, and unit-cell sizes, MCP achieves negligible reconstruction error using tiny data fractions. %
Figure~\ref{fig:compression_comparison} demonstrates that only $\sim$0.1\% of interaction data is sufficient for quantitatively accurate \e-ph couplings in a prototypical ionic solid (LiF), polar semiconductor (GaAs), and 2D material (monolayer MoS$_2$, shown in SM~\cite{supplemental}), and only $\sim$1\% is needed for a correlated insulator (LCO) and a moire system (TBG).
MCP improves compression efficiency over SVD by roughly an order of magnitude across these systems, mirroring the Si benchmark. 
\\
\indent
The gain is most striking in LCO and TBG, where the relatively large unit cells allow coarse phonon momentum sampling: only 8 $\qq$-points for LCO and 9 for TBG.
SVD compresses only the momentum dimension while retaining all internal indices~\cite{SVD-eph-prx2024}, which grow rapidly with system size—in LCO, these correspond to $24^2\times42$ variables associated with 24 Wannier orbitals and 42 phonon modes. Therefore, SVD must retain at least $1/8=12.5\%$ of the original data in LCO and $1/9\approx11.1\%$ in TBG, placing a strict limit on the compression achievable with SVD for such coarse momentum grids.
%
By compressing the internal and momentum dimensions jointly, MCP overcomes this limitation and achieves substantial compression in both systems. %
This makes the tensor representation both necessary and highly effective for materials with large unit cells, where internal degrees of freedom grow rapidly while momentum grids remain small.
\begin{figure*}[t]
    \centering
    \includegraphics[width=1.0\linewidth]{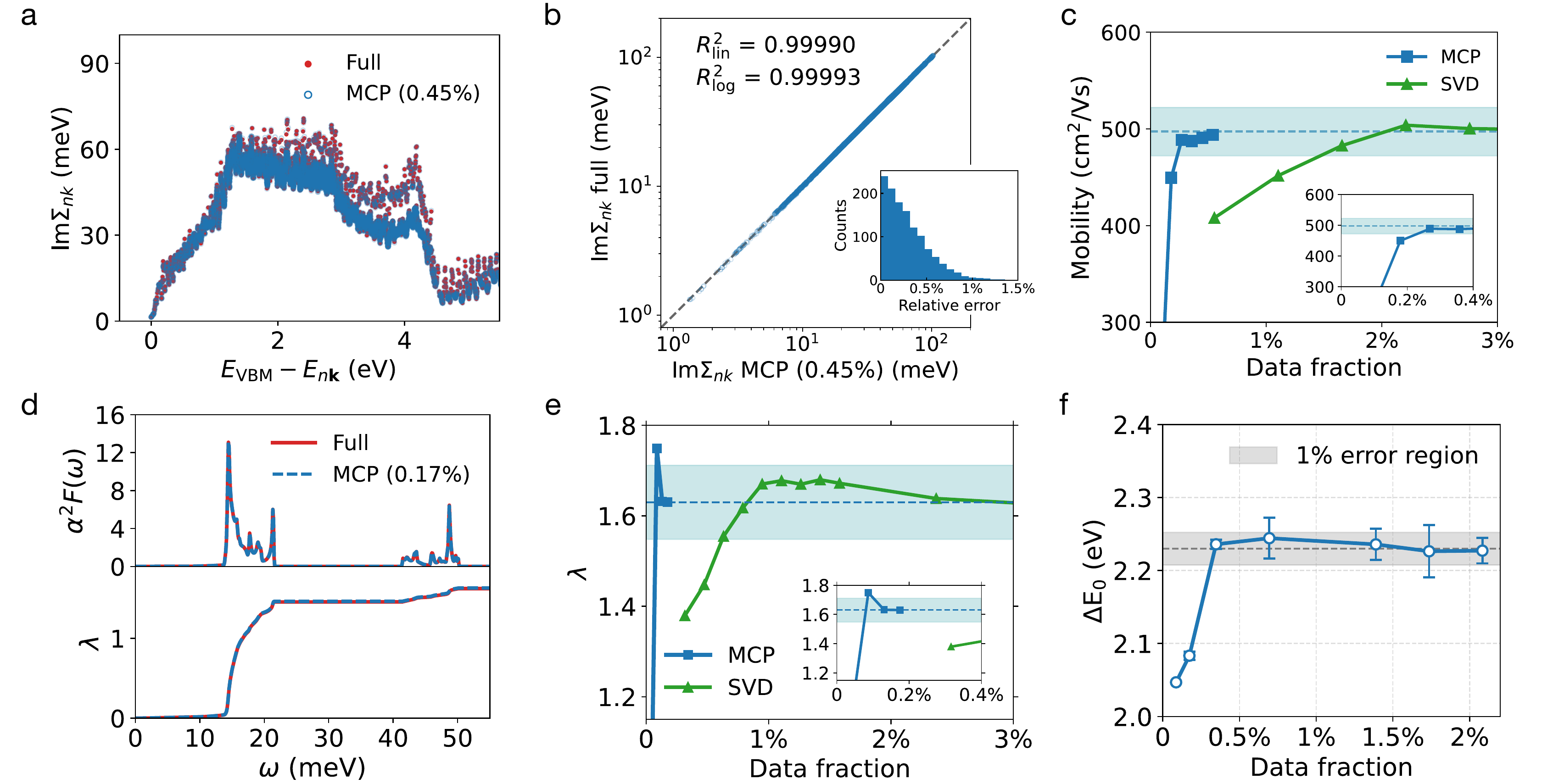}
    \caption{MCP accuracy for \e-ph-derived properties. (a) Si valence-band $\mathrm{Im}\,\Sigma_{n\kk}$ from full and MCP interactions ($p=0.45\%$). (b) Corresponding parity plot; inset: relative-error distribution. (c) Si hole mobility at 300~K within the relaxation-time approximation (RTA) versus $p$. (d) Application to superconductivity, illustrated by showing $\alpha^2F(\omega)$ and cumulative $\lambda(\omega)$ for doped monolayer MoS$_2$ ($p=0.17\%$). (e) Convergence of $\lambda$ with $p$. Panels (c,e) compare MCP and SVD; dashed lines show full-interaction values and shading indicates the $\pm5\%$ windows. (f) LiF hole-polaron formation energy from FEP-DMC versus data fraction $p$; the dashed line and shading show the largest-rank result and its $\pm1\%$ window.}
\label{fig:observables}
\end{figure*}

\subsection*{Application to physical properties}
\vspace{-6pt}
\indent
We next demonstrate the predictive accuracy and computational advantages of MCP-compressed interactions for \e-ph-derived properties, including electronic scattering rates, transport, phonon-mediated superconductivity, and polaron formation energy~\cite{supplemental}.
Figure~\ref{fig:observables}(a)  shows the scattering rates, expressed through the imaginary part of the Fan–Migdal \e-ph self-energy, for Si valence states at 300~K~\cite{Bernardi2016} (see Methods). 
Retaining only $0.45\%$ of the interaction data, MCP recovers with high accuracy the full-interaction scattering rates across a 6-eV energy window. 
The agreement holds for individual electronic states, with coefficients of determination $R_{\mathrm{lin}}^2=0.99990$ on a linear scale and $R_{\mathrm{log}}^2=0.99993$ on a log-scale [Fig.~\ref{fig:observables}(b)]; most rates deviate by less than $1\%$ (see inset) even though their values span two orders of magnitude. 
\\
\indent 
The same compression fidelity extends to transport and superconductivity. For the phonon-limited hole mobility in Si, MCP reaches the $5\%$ accuracy window with only $0.3\%$ of retained data, compared with approximately $2\%$ retained data for SVD [Fig.~\ref{fig:observables}(c)]. 
For phonon-mediated superconductivity in doped monolayer MoS$_2$~\cite{dopedMoS2-prb-2013}, 
at a data fraction of only $p=0.17\%$ MCP recovers both the Eliashberg spectral function $\alpha^2F(\omega)$ and the integrated \e-ph coupling strength $\lambda=2\int_0^\infty d\omega\,\alpha^2F(\omega)/\omega$ with near-perfect agreement with the full interaction result [Fig.~\ref{fig:observables}(d)].
The coupling $\lambda$ converges to the full-interaction value with less than
$0.2\%$ of the data, whereas SVD requires a data fraction ten times larger ~\cite{SVD-eph-prx2024} [Fig.~\ref{fig:observables}(e)]. %
MCP therefore preserves both the frequency-resolved structure and integrated strength of the \e-ph coupling that controls superconductivity.
\\
\indent
We also combine MCP with first-principles diagrammatic Monte Carlo (FEP-DMC) calculations, illustrated here for the case of LiF hole polaron~\cite{luo_first-principles_2025}.
These calculations sample billions of Feynman diagrams to sum \e-ph processes to all orders, requiring repeated evaluation of \e-ph couplings at randomly sampled momenta.
This is computationally prohibitive with conventional Wannier interpolation, making compressed \e-ph representations essential. 
Figure~\ref{fig:observables}(f) shows that MCP yields a polaron formation energy within $1\%$ of the converged result for a data fraction as small as $p\approx0.5\%$. 
%
This rapid convergence demonstrates the efficiency and accuracy of MCP for many-body calculations beyond lowest-order perturbation theory.\vspace{-11pt}

\subsection*{Compression of exciton-phonon interactions}
\begin{figure}[b]
    \centering
    \includegraphics[width=0.75\linewidth]{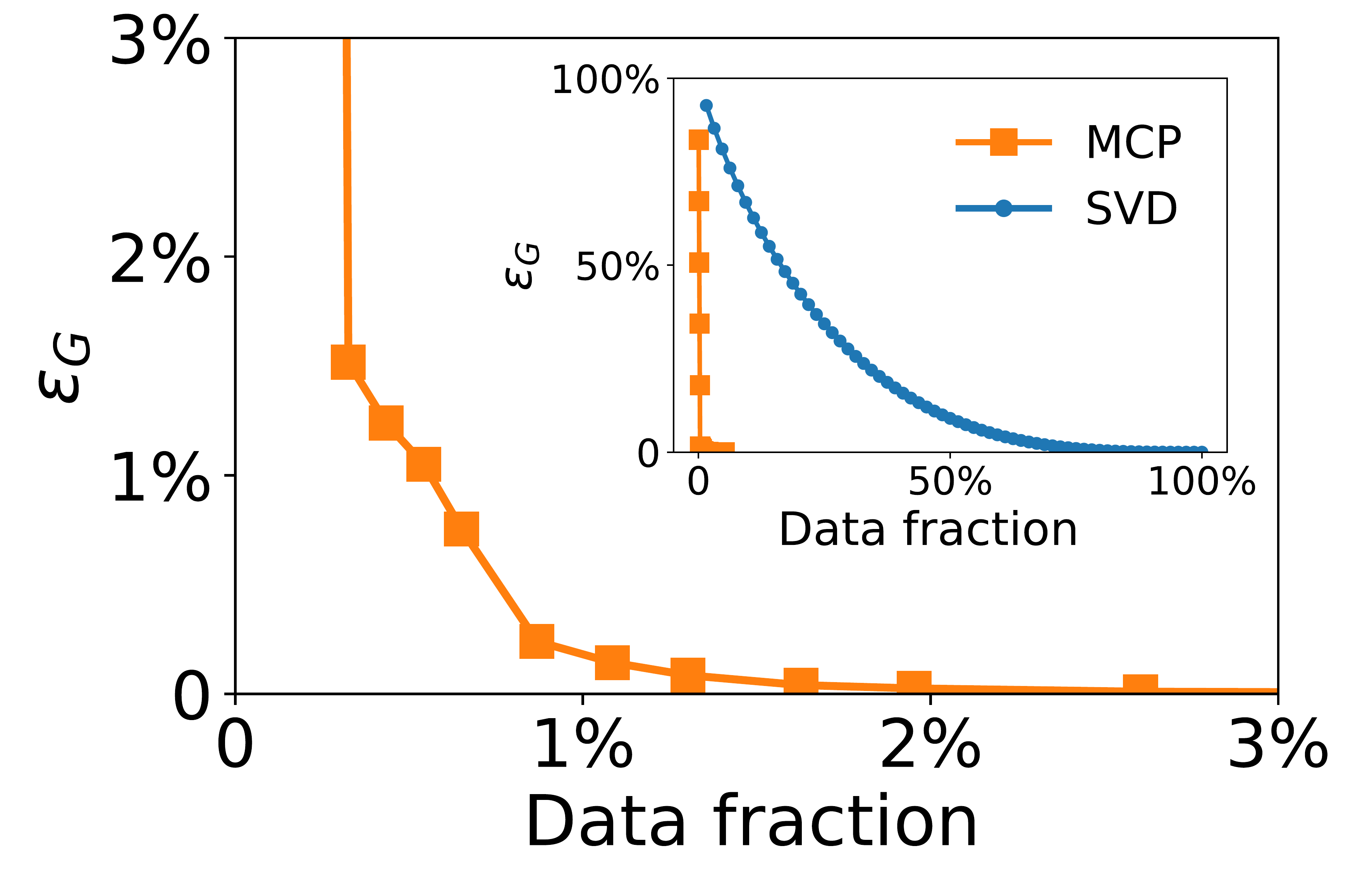}
    \caption{Reconstruction error for exciton-phonon interactions in monolayer CrBr$_3$, comparing MCP and SVD results.}
    \label{fig:exph}
\end{figure}

\vspace{-6pt}
Finally, MCP can also compress interactions involving electronic excitations, which typically lack a smooth gauge representation. 
%
We demonstrate this for first-principles exciton-phonon (ex-ph) interactions,
$G_{SS'\nu}(\mathbf{Q},\qq)$, viewed here as an order-three tensor, where $\mathbf{Q}$ is the exciton center-of-mass momentum and
$S,S'$ label the initial and final exciton states obtained from the Bethe-Salpeter equation (BSE)~\cite{exciton-Rohlfing-prb-2000}.
The interaction has the same three-leg structure as the \e-ph coupling,
so MCP applies directly by replacing the electronic states with BSE exciton states. 
We compress the tensor for the six
lowest exciton bands on $8\times8$ $\mathbf{Q}$- and $\qq$-point grids~\cite{supplemental}.
Figure~\ref{fig:exph} shows that MCP achieves a small reconstruction error,
$\epsilon_G=\|\tilde{\bm{G}}-\bm{G}\|_F^2/\|\bm{G}\|_F^2<1\%$,
using less than $1\%$ of the original data, revealing a hidden low-rank structure for ex-ph interactions in CrBr$_3$.
In contrast, SVD applied to the ex-ph matrices yields negligible compression, requiring most of the singular-value spectrum to achieve comparable accuracy (see inset of Fig.~\ref{fig:exph}). 
This failure has the same origin as for \e-ph SVD in the Bloch gauge: the BSE eigenstates are obtained independently at each $\mathbf{Q}$ point, leading to arbitrary gauge variations across momentum space. MCP absorbs these variations into rotations of its exciton factors, enabling efficient compression without constructing a smooth exciton gauge.

\section*{Discussion}
Beyond accuracy, the key advantage of MCP is its greatly reduced memory footprint and computational cost.
%
For $N_a$ atoms, $N_b$ electronic bands, and $N_{\kk}\!=\!N_{\qq}$ momenta (suppressing the subscript $c$), storing the full \e-ph tensor requires $\mathcal{O}(N_{\kk}^2N_b^2N_a)$ memory, whereas MCP requires only $\mathcal{O}(R_{\mathrm{MCP}}N_{\kk}(N_b+N_a))$, where the rank $R_{\mathrm{MCP}}$ is typically of order 10. MCP therefore reduces the scaling with both momentum grid size and band number from quadratic to linear. 
By contrast, SVD-based compression reduces only the momentum scaling and remains quadratic in the number of bands~\cite{SVD-eph-prx2024}.
%
%
Joint compression of electronic variables makes MCP especially advantageous for materials with
large unit cells, such as LCO and TBG.
The factorized representation also makes interaction evaluations inexpensive: once the CP factors are tabulated, a single \e-ph matrix element
can be assembled through Eq.~\eqref{eq:mcp} using only
$R_{\mathrm{MCP}}\approx\mathcal{O}(10)$ operations, enabling calculations of first-principles \e-ph couplings at a cost comparable to simple model Hamiltonians.
The computational speedup of MCP is proportional to the compression factor $1/p$, directly translating compression into computational time savings (see SM~\cite{supplemental}). 
For instance, in Si and MoS$_2$, MCP achieves accurate predictions ($\varepsilon_g\!<\!1$\%) using only 0.45\% and 0.17\% of the interaction data, respectively, while accelerating interpolation by factors of $\sim$90 and 500~\cite{supplemental}.
\\
\indent
In summary, we establish MCP as an efficient tensor representation that reveals the low-rank structure of \e-ph and ex–ph interactions and substantially accelerates predictive calculations of material properties. 
More broadly, MCP provides a data-driven route to discovering compact representations of microscopic  interactions directly from first-principles data. The framework can be extended to a broad class of couplings, including nonlinear \e-ph and electron–electron interactions, by assigning MCP factors to each fermionic and bosonic degree of freedom and enforcing relevant symmetries. 
Beyond compression, its factorized tensor structure provides natural interfaces with tensor-network methods and quantum algorithms for many-body calculations~\cite{THC-QCC-2021,Jose2026}, as well as with machine learning methods for material discovery~\cite{MACE2022,JornadaDuanXu2026}, where MCP factors could be learned or parametrized with equivariant neural networks to construct transferable models of  interactions across material databases. 
By bringing first-principles interaction models closer to the computational efficiency of model Hamiltonians, MCP opens a path toward data-driven discovery and scalable many-body simulation of complex materials.

\section*{Acknowledgments}
\noindent
\textbf{Funding:}
This work was supported by the Department of Energy under Grant No. DE-SC0026859, which provided for theory, method, and code development. S.S.S. is supported by a Carver Mead Discovery Grant. This research used resources of the National Energy Research Scientific
Computing Center, a DOE Office of Science User Facility supported by the Office of Science of the U.S. Department of Energy under Contract No. DE-AC02-05CH11231 using NERSC award DDR-ERCAP0036498.

\bibliography{ref}

\end{document}